\documentclass[twocolumn,superscriptaddress,floatfix,preprintnumbers, nofootinbib,hyperref]{revtex4} 
\pdfoutput=1
\usepackage[colorlinks=true,breaklinks=true]{hyperref}
\usepackage[normalem]{ulem}
\usepackage[utf8]{inputenc}
\hypersetup{allcolors=[rgb]{0.0 0.0 0.6},linkcolor=[rgb]{0.75 0.05 0.05}}
\usepackage{amsmath,amssymb}
\usepackage{epsfig}  
\usepackage{graphicx}   
\usepackage{slashed}             
\usepackage{url}
\usepackage{color}
\usepackage{multirow}
\usepackage[dvipsnames]{xcolor}
\usepackage{letltxmacro}
\LetLtxMacro{\oldcite}{\cite}
\renewcommand{\cite}[1]{\mbox{\oldcite{#1}}}

\DeclareMathOperator{\MeV}{MeV}

\newcommand{\bk}{{\bf k}}

\newcommand{\beq}{\begin{equation}}
\newcommand{\eeq}{\end{equation}}

\allowdisplaybreaks

\bibpunct{[}{]}{,}{n}{}{,}

\begin{document}

\title{
Axion-like Particles from Hypernovae
}

\author{Andrea Caputo}
\affiliation{School  of  Physics  and  Astronomy,  Tel-Aviv  University,  Tel-Aviv  69978,  Israel}
\affiliation{Department  of  Particle  Physics  and  Astrophysics,Weizmann  Institute  of  Science,  Rehovot  7610001,  Israel}%
\affiliation{Max-Planck-Institut f\"ur Physik (Werner-Heisenberg-Institut), F\"ohringer Ring 6, 80805 M\"unchen, Germany} 

\author{Pierluca Carenza}
\affiliation{Dipartimento Interateneo di Fisica “Michelangelo Merlin”, Via Amendola 173, 70126 Bari, Italy}
\affiliation{Istituto Nazionale di Fisica Nucleare - Sezione di Bari, Via Orabona 4, 70126 Bari, Italy}%

\author{Giuseppe Lucente}
\affiliation{Dipartimento Interateneo di Fisica “Michelangelo Merlin”, Via Amendola 173, 70126 Bari, Italy}
\affiliation{Istituto Nazionale di Fisica Nucleare - Sezione di Bari, Via Orabona 4, 70126 Bari, Italy}%

\author{Edoardo Vitagliano}
\affiliation{ Department  of  Physics  and  Astronomy,  University  of  California,  Los  Angeles,  California,  90095-1547,  USA}

\author{Maurizio Giannotti}
\affiliation{Physical  Sciences,  Barry  University,  11300  NE  2nd  Ave.,  Miami  Shores,  FL  33161,  USA}%

\author{Kei Kotake}
\affiliation{Department of Applied Physics \& Research Institute of Stellar Explosive Phenomena, Fukuoka University, Fukuoka 814-0180, Japan}%

\author{Takami Kuroda}
\affiliation{Max-Planck-Institut f\"ur Gravitationsphysik, Am M\"uhlenberg 1, D-14476 Potsdam-Golm, Germany}%

\author{Alessandro Mirizzi}
\affiliation{Dipartimento Interateneo di Fisica “Michelangelo Merlin”, Via Amendola 173, 70126 Bari, Italy}
\affiliation{Istituto Nazionale di Fisica Nucleare - Sezione di Bari, Via Orabona 4, 70126 Bari, Italy}%

\date{\today}
\smallskip


\begin{abstract}
It was recently pointed out that very energetic subclasses of supernovae (SNe), like hypernovae and superluminous SNe, might host ultra-strong magnetic fields in their core. 
Such fields may catalyze the production of feebly interacting particles, changing the predicted emission rates.
Here we consider the case of axion-like particles (ALPs) and show that the predicted large scale magnetic fields in the core contribute significantly to the ALP production, via a coherent conversion of thermal photons. 
Using recent state-of-the-art SN simulations including magnetohydrodynamics, we find that if ALPs have masses $m_a \sim {\mathcal O}(10)\, \rm MeV$, their emissivity via magnetic conversions is over two orders of magnitude 
larger than previously estimated. Moreover, the
radiative decay of these 
massive ALPs
 would lead to a
peculiar delay in the arrival times of the daughter 
photons. Therefore, high-statistics gamma-ray satellites can potentially discover  MeV ALPs in an unprobed region of the parameter space and shed light on the magnetohydrodinamical nature of the SN explosion. 
 \end{abstract}

\maketitle

\textit{Introduction---}Core collapse supernovae (SNe) are recognized as  extremely efficient laboratories  for light, feebly interacting particles, 
such as neutrinos, axion-like particles (ALPs), and dark photons~\cite{Raffelt:1999tx,Raffelt:1996wa}.
Very often, SN bounds significantly constrain the viable parameter space for these particles, providing guidance for their experimental searches. 
The enthusiasm for SNe as laboratories for astroparticle physics has grown in recent years, thanks to advancements in numerical simulations
and the development of new neutrino detectors, which might provide important information about both the explosion mechanism of SNe  and fundamental physics (see e.g. Refs.~\cite{Mirizzi:2015eza,Horiuchi:2017sku,Agrawal:2021dbo} for recent reviews).

The current paradigm of the SN explosion mechanism is based on the neutrino-driven scenario, in which neutrino energy deposition
revitalizes the stalled shock-wave (see e.g. Refs.~\cite{Janka:2016fox,Janka:2017vcp}).
However,  there are some very energetic subclasses of supernovae, including hypernovae and superluminous SNe~\cite{Moriya}, which are highly unlikely to be explained by the conventional
 neutrino-driven explosion.  The most plausible scenario to account for these extreme events requires additional energy injection via the magnetohydrodynamically-driven (MHD) explosions~\cite{Burrows:2007yx}.
 This situation has been explored in very recent dedicated numerical studies~\cite{Matsumoto:2020rbz,Obergaulinger:2020cqq},
 finding that in this case the SN core might host ultra-strong magnetic fields ($B \gtrsim 10^{15}\, \rm G$~\cite{Mosta:2015ucs}).

In light of these recent developments, it becomes essential to consider the question of whether the presence of potentially 
very large magnetic fields might influence significantly the emission of light, feebly interacting particles 
from SNe.
It is known, for example, that if neutrinos possess a magnetic moment $\mu \gtrsim 10^{-13} \mu_B$, where $\mu_B$ is the 
Bohr magneton,
strong magnetic fields may affect
their flavor conversions in the SN core, leading to peculiar observational signatures 
(e.g.~\cite{Akhmedov:2003fu,Abbar:2020ggq}).

However, the impact on other feebly interacting particles has never been studied before. 
Here we consider ALPs, 
which are among the most prominent and well studied new physics candidates~\cite{Ringwald:2014vqa,Agrawal:2021dbo},
 and show that the presence of a magnetic field may have a 
 significant impact on the emission of these particles 
 if their mass falls in the  $m_a \sim {\mathcal O}(10)\, \rm MeV$ mass range. \\
\textit{Axion-like particles and supernovae---}ALPs are pseudoscalar bosons $a$ with a two-photon vertex, described by the
 Lagrangian~\cite{Sikivie:1983ip}
\begin{equation}
{\mathcal L}_{a \gamma} = -\frac{1}{4} g_{a\gamma} F_{\mu \nu} {\tilde F}^{\mu \nu}a  \,\ ,
\end{equation}
where $g_{a\gamma}$ is the effective ALP-photon vertex,  $ F_{\mu \nu}$ is the electromagnetic field and ${\tilde F}^{\mu \nu}$ its dual.
These particles emerge 
in different extensions of the Standard Model (see Refs.~\cite{Agrawal:2021dbo,DiLuzio:2020wdo} for recent reviews) and are object of an intense experimental investigation~\cite{Irastorza:2018dyq,Sikivie:2020zpn}. 
Furthermore, astrophysical observations on different energy scales offer valuable 
opportunities to search for these particles~\cite{Raffelt:2006cw,
Giannotti:2017hny,Giannotti:2015kwo}.

Core-collapse SNe are particularly powerful cosmic factories 
for ALPs~\cite{Raffelt:1996wa,Raffelt:2006cw}.
Notably, the SN 1987A neutrino detection has been a milestone event also for axion physics~\cite{Turner:1987by,Burrows:1988ah,Lucente:2020whw}. 
 The dominant production channel  in SNe for light ALPs coupling exclusively to photons is the 
 Primakoff emission~\cite{Raffelt:1985nk}, in which thermal photons are converted into ALPs in the electrostatic field of protons.
Once produced, SN ALPs provide different signatures depending on their mass. Very light ALPs ($m_a \lesssim 10^{-10}\,\rm eV$) 
leave the star and convert into gamma rays in the 
magnetic field of the Milky Way~\cite{Grifols:1996id,Brockway:1996yr}. 
Indeed, the lack of a gamma-ray signal in the
 Gamma-Ray Spectrometer (GRS) on the Solar Maximum Mission (SMM) in coincidence with the observation of the neutrinos emitted from SN 1987A provides a strong bound on the ALP coupling to photons~\cite{Grifols:1996id,Brockway:1996yr}.
The most recent analysis finds
 $g_{a\gamma} < 5.3 \times 10^{-12}\,\rm GeV^{-1}$, for $m_a < 4 \times 10^{-10}\, \rm eV$~\cite{Payez:2014xsa}.
On the other hand, ALPs with $m_a\sim {\mathcal O}(0.1-100)\,\rm MeV$   decay into photons, thus producing a gamma ray flux without the need for a galactic magnetic field.
In this case, a similar analysis gives 
 $g_{a\gamma} \lesssim   10^{-11}\,\rm GeV^{-1}$ at $m_{a}\simeq 10\,\rm MeV$~\cite{Jaeckel:2017tud}.
 Intriguing opportunities to sharpen these bounds 
 are offered by the detection of an ALP burst 
 in future (extra)-galactic SN explosions~\cite{Meyer:2016wrm,Meyer:2020vzy}
  or from 
 the diffuse ALP flux from all past core-collapse SNe in the Universe~\cite{Calore:2020tjw}.
 If detected, the ALP flux would carry information on the deepest SN regions,
 complementary to the ones provided by neutrinos and gravitational waves~\cite{Nakamura:2016kkl,Li:2020ujl}.

Magnetic fields could play a significant role in the ALP production itself. ALPs may be produced by the scattering of photons on an external macroscopic magnetic field, rather than on the electrostatic field of charged particles as in Primakoff emission.
Recent investigations have shown, for example, that the magnetic ALP production in the Sun may dominate over the Primakoff production, particularly in some energy ranges~\cite{Caputo:2020quz,OHare:2020wum,Guarini:2020hps}.
An equivalent analysis for the SNe is lacking because of the absence, until very recently, of numerical SNe simulations which included also a magnetic field. 
In this work we fill this gap and investigate the role that ultra-strong magnetic fields in SN core have in the emissivity of ALPs. 

\textit{ MHD Supernova model---}
For this purpose we consider a fully relativistic 3D magnetorotational core-collapse SN simulation with multi-energy neutrino transport~\cite{Kuroda:2020pta}.
The progenitor model is a $20 \,M_{\odot}$ star with solar-metallicity from Ref.~\cite{Woosley:2007as},
while the nuclear part is described by the equation of state SFHo~\cite{SFH}.
Since the original progenitor model assumed neither rotation nor magnetic field during its evolution, a cylindrical rotation and a dipolar-like magnetic field was added in Ref.~\cite{Kuroda:2020pta}. 
The resulting star has a central initial angular frequency and magnetic field strength of $1$ rad s$^{-1}$ and $10^{12}\, \rm G$ respectively. These lead to a dynamically relevant magnetic field of $B\gtrsim 10^{15}\, \rm G$ inside the proto-neutron star after bounce through the magnetic compression and winding effects.
This model experiences the so-called magnetorotational explosion soon after bounce and the shock wave reaches $\sim4000$ km from the center at the final simulation post-bounce time~$t_{\rm pb}=370$ ms.

\begin{figure}[t]
\centering
\includegraphics[width=0.48\textwidth]{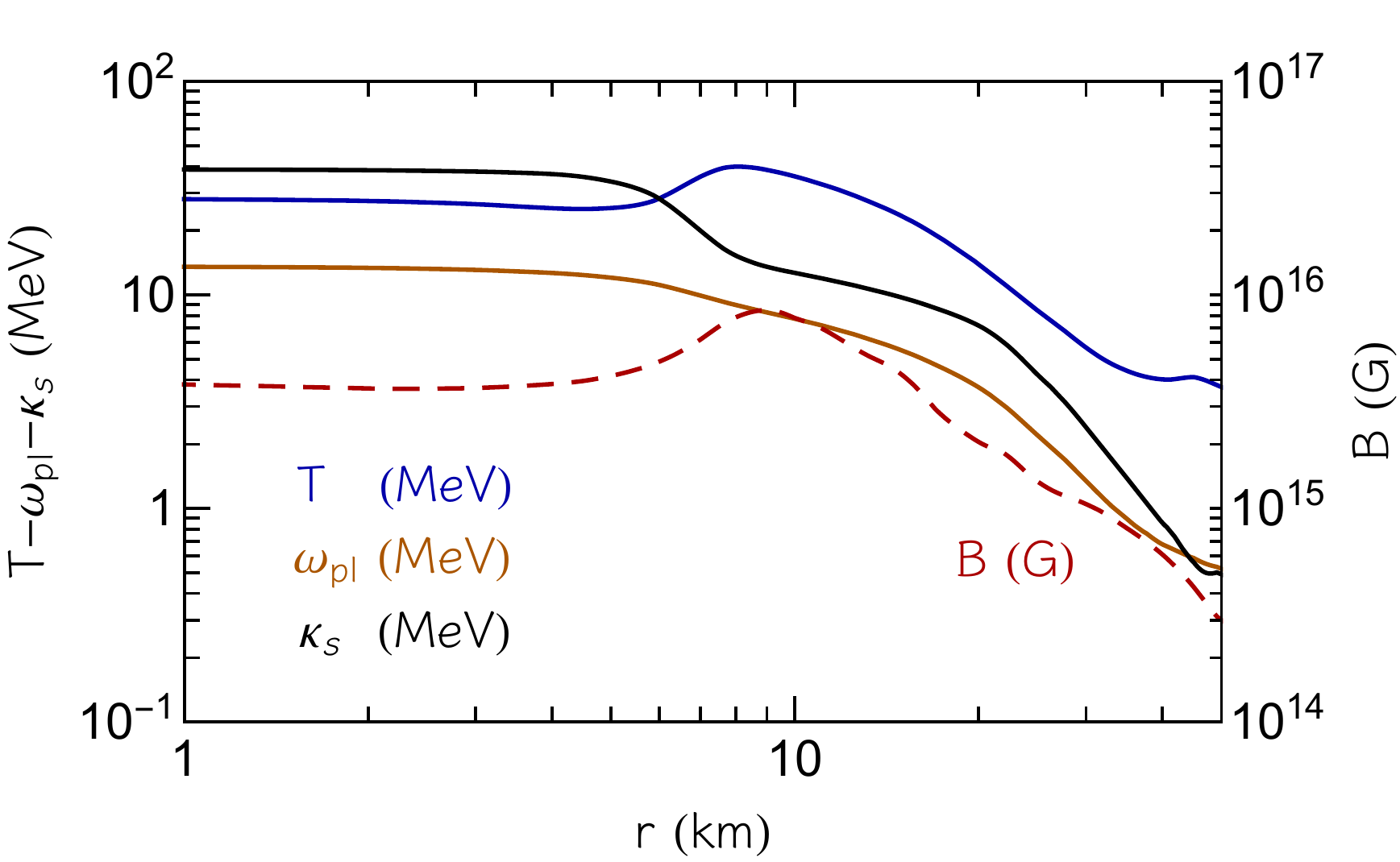}
\caption{Angle-averaged radial profile for the temperature $T$ (blue), the plasma frequency $\omega_{\rm pl}$ (orange), the screening scale $\kappa_s$ (black) and the magnetic field strength $B$ (dashed red) at $t_{\rm pb}=370\,\rm ms$.}
\label{fig:pl_T_B_profile}
\end{figure}

In Fig.~\ref{fig:pl_T_B_profile}, we show the angle-averaged radial profile of the magnetic field strength $B$ (dashed red) at ${t_{\rm pb}=370\, \rm ms}$ in the SN core ($r\lesssim 50\,\rm km$) for the SN model we are considering. 
In the very inner core $r\lesssim5\,\rm km$ the magnetic field assumes a constant value $B\simeq 4\times 10^{15}\, \rm G$. It reaches its maximum $B_{\text{max}}\simeq 7\times 10^{15}\, \rm G$ at $r\simeq 10\,\rm km$, where the strong shear motion appears, while it decreases at larger radii, becoming smaller than $10^{15}\, \rm G$ at $r\gtrsim30\, \rm km$. 

We also show the angle-averaged radial profiles for the
temperature $T$ (blue), the plasma frequency $\omega_{\rm pl}$
(orange), the screening scale $\kappa_s$ (black) in the inner SN core. 
The plasma frequency, defined as
$\omega_{\rm pl} \simeq 16.3\,\text{MeV}\,(Y_e\rho_{14})^{1/3}$ \cite{Kopf:1997mv}, where  $\rho_{14}=\rho/10^{14}\,\text{g cm}^{-3}$ and $Y_e$ is the electron fraction, plays the role of an ``effective photon mass''. 
The screening scale
 $\kappa_s$ accounts for the correlation effects of the charged particles in the 
 stellar medium. 
 Since protons are only partially degenerate, a good estimate of 
 the screening scale is given by $\kappa_s^2=4\pi\alpha\,n/T$ (see  discussion in Ref.~\cite{Payez:2014xsa} for the effects of partial degeneracy). 

\textit{ ALP Primakoff emission---}The main ALP production channel in a SN core for ALPs with masses  $m_{a}\lesssim80\,\MeV$  is the Primakoff process. Since electrons are highly degenerate in the SN core, their phase space is Pauli blocked and their contribution to the ALP production is negligible. Therefore, the most substantial contribution to the Primakoff process comes from free protons, which are only partially degenerate in the SN core. 
With these assumptions, the Primakoff production rate for massive ALPs is~\cite{DiLella:2000dn,Lucente:2020whw}
\begin{align}
\Gamma_P&=2\,g_{a\gamma}^2\dfrac{T\kappa_s^2}{32\pi} \dfrac{p}{E}\bigg\{\dfrac{\left[\left(k+p\right)^2+\kappa_s^2\right]\left[\left(k-p\right)^2+\kappa_s^2\right]}{4kp\kappa_s^2} \nonumber \\*
\times &\ln\left[\dfrac{(k+p)^2+\kappa_s^2}{(k-p)^2+\kappa_s^2}\right]
-\dfrac{\left(k^2-p^2\right)^2}{4kp\kappa_s^2}\ln\left[\dfrac{(k+p)^2}{(k-p)^2}\right]-1\bigg\}\,, \nonumber \\*
\label{eq:Prima}
\end{align}
where the factor $2$ accounts for the photon polarization states, $E$ and $p$ are respectively the energy and momentum of the ALP, and $k$ is the momentum of the photon.
In the conditions of interest for us, we can neglect the proton recoil energy and take the photon energy equal to the ALP energy, $\omega=E$.

\textit{ALP emission via magnetic conversions---}
The rate for photon-ALP conversion in a magnetized plasma was recently 
calculated in Refs.~\cite{Caputo:2020quz,OHare:2020wum,Guarini:2020hps}.
ALPs can be resonantly produced in an external $B$ field when the dispersion relations of longitudal ($L$) or transverse ($T$) photon degrees of freedom match the dispersion relation of the ALP.
The resonant emission rate is~\cite{ Mikheev:1998bg, Caputo:2020quz,Guarini:2020hps} 
\begin{align}
 \Gamma_{B}=&\,m_a^2B_{||}^2g_{a\gamma}^2\frac{\pi}{2\omega^2}Z_L\delta(\omega-\omega_0^L(k))  \nonumber \\*
	+&\,\omega^2B_{\perp}^2g_{a\gamma}^2\frac{\pi}{2\omega^2}Z_T\delta(\omega-\omega_0^T(k))\,, \label{eq:imPIaxion}
\end{align}
where the factor $Z_{L,T}$ can be interpreted as renormalizing the coupling to the axion, $B_{||}$ and $B_{\perp}$ are the $B$ field components parallel and orthogonal to the photon momentum, and $\omega_0^{T,L}(k)$ is the dispersion relation for the considered electromagnetic mode~\cite{Raffelt:1996wa} (see the Appendix for further details). 

A word of caution is in order. The presence of very strong magnetic fields, $B \simeq 10^{16}\, \rm G$, 
could potentially affect the electron wave-functions and,
moreover, generate QED nonlinear effects. 
However, SNe feature a strongly degenerate and ultra-relativistic plasma, so the Fermi momentum is much larger than the synchrotron frequency $\omega_B=eB/p_F$. 
In such environments, the gas behaves quasi-classically, as the magnetic field does not quantize the electron wave-functions~\cite{kaminker1991,Bezchastnov:1997ew,Yakovlev:2000jp}. This hierarchy also prevents from the production of 
electron-positron pairs.
Moreover, the electron mass is modified to $m_e^2 \simeq \frac{e^2}{2\pi^2} p_F^2 \gg (m_e^{\text{vac}})^{2}$~\cite{Bellac:2011kqa,PhysRevD.31.3280}, which implies that the critical magnetic field $B_c=m_e^2/e$ is much larger than in vacuum. 
Thus, we can neglect also birefringence 
effects~\cite{Tsai:1974fa},\footnote{Note that for $B \gtrsim B_c$ the correct equations for vacuum birefringence are given in Ref.~\cite{Tsai:1974fa} and differ from those commonly found in the literature~\cite{Raffelt:1996wa}, where the implicit assumption is $B \ll B_c$.} which typically inhibit the conversion in intense magnetic fields in vacuum~\cite{Raffelt:1987im}. 
Therefore, we identify a previously overlooked condition for efficient ALP-photon conversion in exceptionally strong magnetic fields which does not rely on the ALP being nonrelativistic~\cite{Hook:2018iia}.

\begin{figure}
 \includegraphics[width=0.45\textwidth]{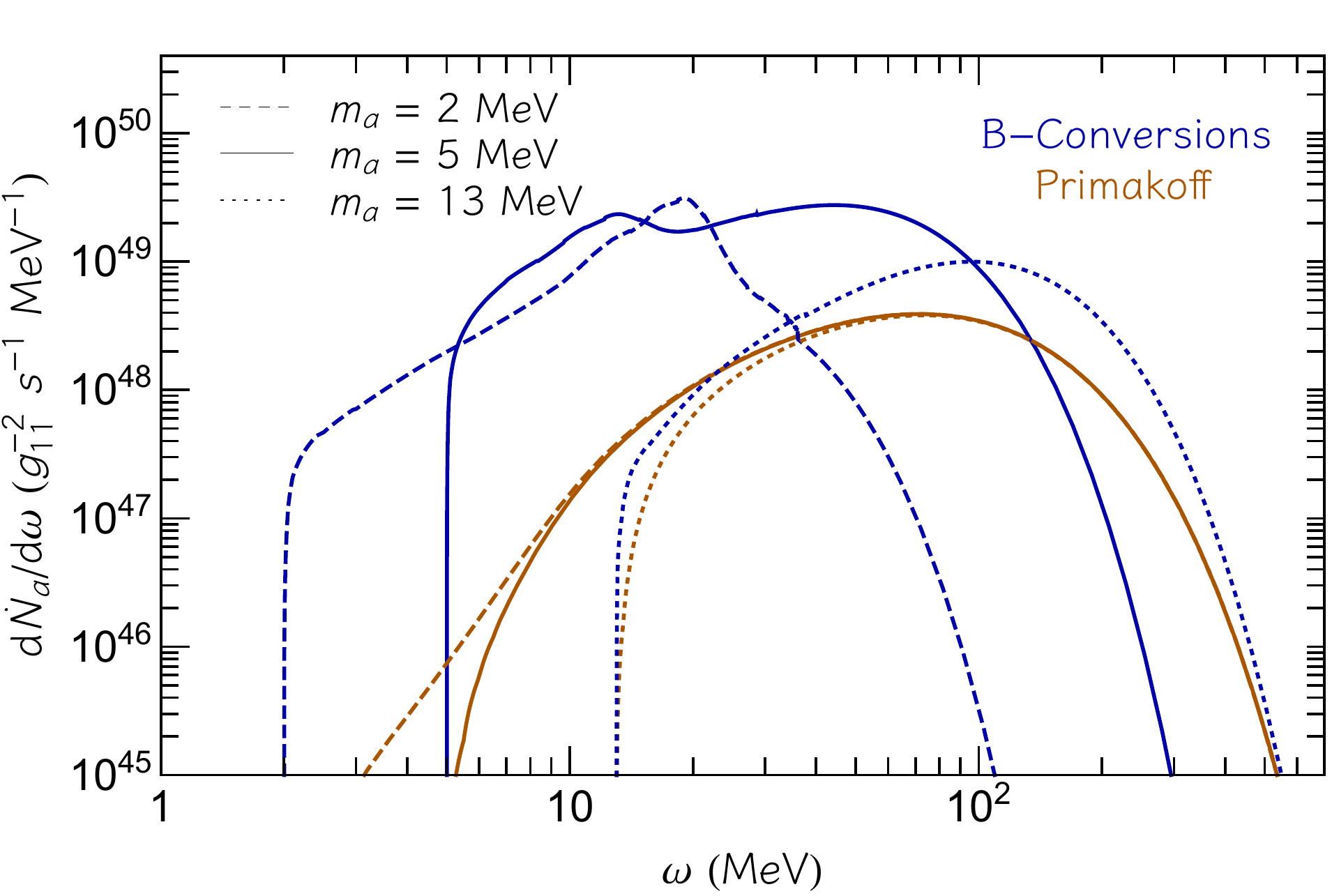}
\caption{Differential ALP production rate for the Primakoff  (orange) and the $B$-conversion (blue) processes, for $m_a =2, 5, 13\, \rm MeV$ (dashed, solid and dotted lines respectively) and $g_{a\gamma}=10^{-11}\, \rm G$eV$^{-1}$. In the case of $m_a = 2, 5\,\rm MeV$, at
$\omega\gtrsim 10\, \rm MeV$, we notice a small peak due to the presence---in a very narrow energy range---of longitudinal modes conversion.
}
\label{fig:flux}
\end{figure}

\textit{Comparison of ALP fluxes---}
The ALP emissivity for a given process is found by integrating  the production rate over the photon thermal spectrum
\begin{equation}
    Q_a=\int\frac{d^{3}\bk}{(2\pi)^{3}} \frac{\Gamma\omega}{e^{\omega/T}-1}\,,
   \label{Qa}
\end{equation}
and the ALP luminosity by integrating the previous expression over the SN model,
\begin{equation}
L_a= \int d\Omega \int Q_a(r,\Omega) r^2 dr \,\ = \int d\omega\, \omega \frac{dN_a}{d\omega\, dt}\, ,
\label{eq:lumen}
\end{equation}
where $dN_a/d\omega dt$ is the differential ALP production rate. The latter is shown in Fig.~\ref{fig:flux} for three different values of the ALP mass, 
$m_a = 2, 5, 13\, \rm MeV$. 
The plasmon-ALP conversion clearly dominates over Primakoff, especially for energies $\omega \lesssim 200 \, \rm MeV$, 
if the ALP mass is in the range $ 4 \lesssim m_a \lesssim  14\, \rm MeV$. 
For $m_a \lesssim 4\, \rm MeV$, the resonance conversion takes place in the  external regions of the star, where the magnetic field is too weak to lead to a sizable production.
On the other hand, if the mass is too large, $m_a \gg 14\, \rm MeV$, the conversion happens in the very inner core ($r\lesssim 5\,\rm km$), where the plasmon population is suppressed. 
Notice that the contribution of the $L$-mode is relevant only in a very narrow energy range, at low energy.
At large energies, which---as we shall see below---are the most interesting for a possible detection, the transverse are the only relevant modes. 
Integrating the differential flux gives the total luminosity. 
We find that the B-conversion luminosity is larger than the Primakoff luminosity, $L_{a,B} > L_{a,P}$, for $ 4\,\mathrm{MeV} \lesssim m_a \lesssim 14\, \rm{MeV}$, with a peak luminosity ${L_{a,B}^{\rm max}\simeq 10^2 \,L_{a,P} \simeq 10^{49} \left(\frac{g_{a\gamma}}{10^{-11}\mathrm{GeV}^{-1}}\right)^2\,\rm erg\, s^{-1}}$ for $m_a \simeq 10\, \rm MeV$.

 After being produced in the SN core, ALPs with $\mathcal{O}(1)\, \rm MeV$ mass can decay into two photons with energy
$E_{\gamma}= \omega/2$,
 producing a large gamma-ray flux that can be observed with existent~\cite{Atwood:2009ez} and planned~\cite{DeAngelis:2017gra,McEnery:2019tcm} gamma-ray detectors. 
 This flux will feature a characteristic time distribution, as the distance travelled by the ALP before decaying plus the distance travelled by daughter photons is larger than the distance of the SN from Earth.

Following Refs.~\cite{Giannotti:2010ty,Jaeckel:2017tud}, we estimate the time delay of an ALP-originated photon arriving at Earth with respect to a massless particle coming directly from the SN, as
\begin{equation}
t= \frac{L_1}{\beta}+L_2-d_{SN}\,,
\label{deltat}
\end{equation}
where $\beta$ is the ALP velocity, $L_1=
{64~\omega  \pi}/({g_{a\gamma}^2\,m_a^4)} \times
\sqrt{1-\left({m_a}/{\omega}\right)^2}
$ is the ALP decay length, and  $L_2 = -L_1\cos\alpha + \sqrt{d_{SN}^2-L_1^2\sin^2\alpha}$, with $\alpha$  
the angle between the photon and the ALP momentum, 
is the distance covered by the daughter photon before reaching Earth.
In the relativistic limit,
$\sin\alpha \simeq (\omega/m_a)^{-1} $~\cite{Jaeckel:2017tud}.
Finally, $d_{SN} $ is the Earth-SN distance, 
which we fix to $d_{SN}=10\,\rm kpc$
and, in all cases of interest for us, is much larger than $L_1$. 

The flux of daughter photons can be observed over the interval of several days with gamma ray instruments such as the Fermi Large Area Telescope (LAT)~\cite{Atwood:2009ez}.
In general, the number of events per unit time from the daughter photons expected in a gamma ray detector with a delay time $\tau$ can be estimated as:
\begin{equation}\label{EQ:events}
\begin{split}
\dot{N}_\gamma(\tau)=\frac{2}{4\pi d_{SN}^2}&\left(\frac{d t(\omega^{*}_{\tau})}{d\omega}\right)^{-1} \frac{dN_a(\omega^{*}_{\tau})}{d\omega} A(\omega^{*}_{\tau}/2) \\ &\times\, e^{-R^*/L_1} \,(1-e^{-d_{SN}/L_1})\,,
\end{split}
\end{equation}
where $dN_a/d\omega$ is the SN ALP production rate, $dt/d\omega$ is obtained from Eq.~\eqref{deltat}, $R^*=10^{14}\,\rm cm$~\cite{DeRocco:2019njg} is the effective SN radius and 
the factor $e^{-R^*/L_1}$ ensures that only ALPs decaying outside the SN are counted.  
Analogously, the factor $(1-e^{-d_{SN}/L_1})$ selects ALPs decaying before reaching Earth, and $A(\omega)$ is the detector effective area. Notice that Eq.~\eqref{EQ:events} should be evaluated at $\omega=\omega^{*}_{\tau}$,
where the ALP energy $\omega^{*}_{\tau}$ is obtained by solving Eq.~\eqref{deltat} with $t=\tau$. 

\begin{figure}[t!]
	\centering
	\includegraphics[width=0.45\textwidth]{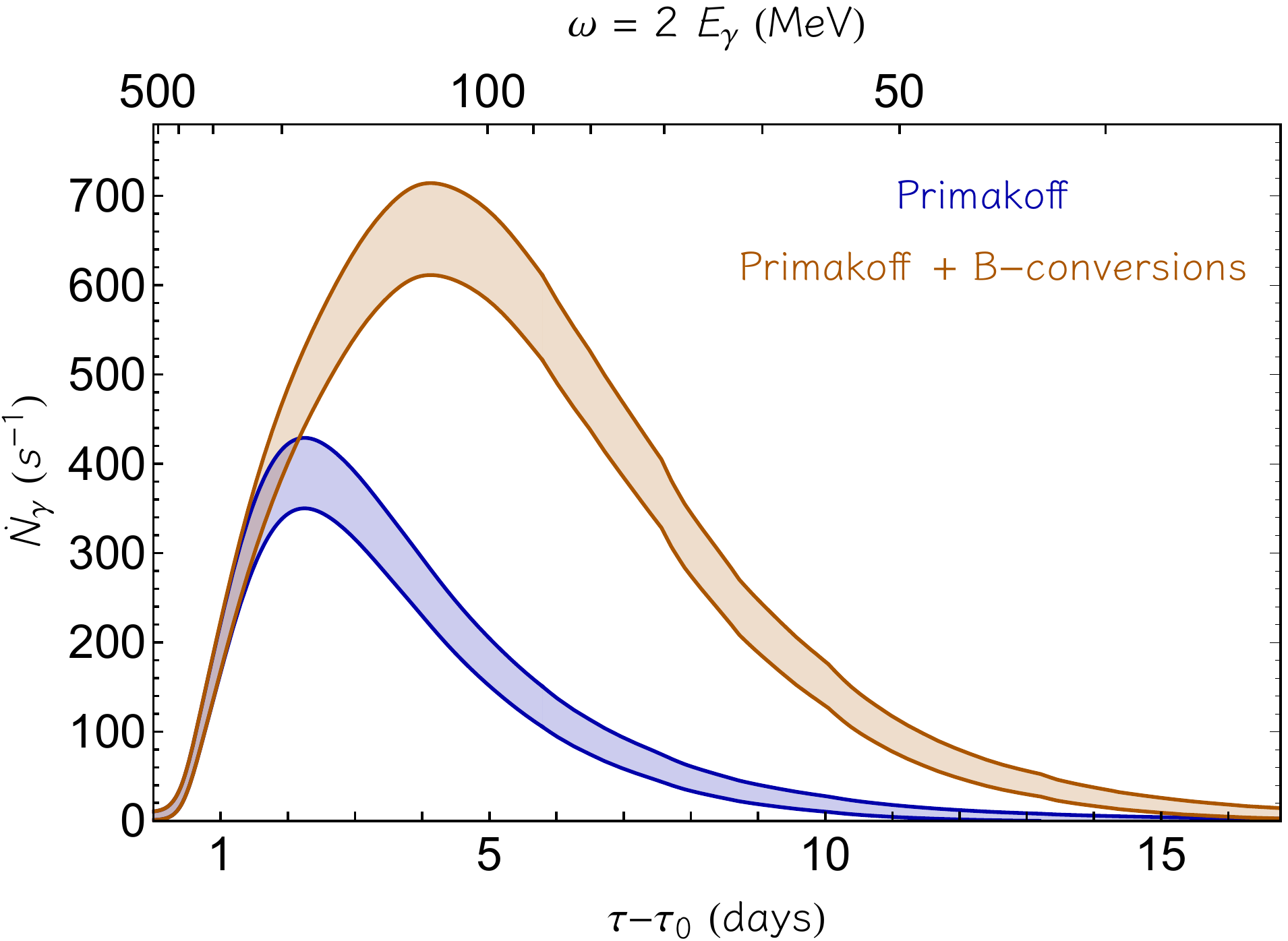}
	\caption{
	Number of events per unit time with $2\sigma$ Poissonian fluctuation contours expected to be observed by Fermi-LAT, as a function of the time delay, obtained considering only Primakoff (blue), and both Primakoff and B-conversions (orange), for $m_a=5\,\rm MeV$ and $g_{a\gamma}=10^{-11}\, \rm G$eV$^{-1}$. $\tau_0=10^5\,\rm s$ is the time-delay of photons produced by ALPs with energy $\omega\simeq 500\,\rm MeV$.}
	\label{fig:delay}
\end{figure}

Here, to give a realistic assessment of the potential to detect photons from ALP decay, we consider the specific case of Fermi LAT with Pass 8 event selection~\cite{Bruel:2018lac}.\footnote{\url{https://www.slac.stanford.edu/exp/glast/groups/canda/lat_Performance_files}}
In Fig.~\ref{fig:delay}, we show the number of expected
events per unit time $\dot{N}_\gamma$---with $2\sigma$ Poissonian fluctuation contours---as a function 
of the time delay $\tau-\tau_0$, where
$\tau_0=10^5\,\rm s$ is a reference time-delay corresponding to photons produced by ALPs with energy $\omega\simeq 500\,\rm MeV$. 
The blue band refers to photons from the decay of Primakoff ALPs, and the orange band, to photons from ALPs produced in the SN magnetic field.
In both cases, we are fixing the ALP mass to 
$m_a=5\,\rm MeV$ and the ALP-photon coupling to $g_{a\gamma}=10^{-11}\, \rm G$eV$^{-1}$.

Though the two signals have a qualitatively  similar 
time evolution, the difference between them is manifest.
Since ALPs produced through magnetic conversion are considerably less energetic than Primakoff ALPs, 
their daughter photons arrive with a more substantial delay. 
Thus, the photon flux in the first couple of days, being
dominated by Primakoff ALPs, has a very weak dependence on the magnetic field,  
and would allow to fix the ALP coupling to photons. 
The photon flux of the following days, after the Primakoff contribution decline, could then provide 
information about the intensity of the magnetic field.
Therefore, reconstructing the time evolution of the signal allows
to gain confidence about the presence of a strong magnetic field in the SN, and to measure its intensity.

\textit{ Conclusions---} Very energetic supernovae, such as hypernovae and superluminous SNe, 
host intense magnetic fields in their cores. 
We have shown that these fields would trigger a resonant ALP production via conversion of photons. Furthermore, we have demonstrated that if ALPs have a mass in the range $ 4\,\rm MeV \lesssim m_a \lesssim 14\, \rm MeV$, the B-conversion would overcome the Primakoff production by over an order of magnitude, becoming the dominant ALP production mechanism in SNe with $B \sim {\mathcal O}(10^{15})\, \rm G$. 

These massive ALPs, with  $m_a\sim {\mathcal O}(10)\, \rm MeV$ and $g_{a\gamma}\sim 10^{-11}\, \rm G$eV$^{-1}$,
have a decay length much shorter than the SN distance and so would produce an observable photon flux on Earth. Depending on the axion coupling, the flux might be observable by gamma ray instruments such as Fermi LAT for well over a week. 
Furthermore, the temporal distribution of the flux is peculiar of the production mechanisms, providing a transparent way to recognize if ALPs are produced through conversions in a magnetic field or through the Primakoff process. 
Finally, since the magnetic production mechanism is very
sensitive to the ALP mass, an accurate analysis of the 
time delay would allow to infer $m_a$.

The discovery of a gamma ray flux from a galactic SN with the temporal properties described in this work would have enormous consequences both for ALP and SN physics. 
On the one hand, it would point out the existence of ALPs and allow to pin down its mass, 
on the other, it would be the smoking gun for the presence of ultra-strong magnetic fields in the SN core. 
This confirms once more 
SNe as multi-messenger laboratories for particle physics and shows how much could be learned about SNe 
from the detection of   
ALP-induced processes.

\bigskip
\textit{Acknowledgments---} 
We warmly thank Niccolò Di Lalla, Georg Raffelt and Jin Matsumoto for useful discussions. 

AC acknowledges support from the the Israel Science Foundation (Grant No. 1302/19), the US-Israeli BSF (grant 2018236) and the German Israeli GIF (grant I-2524-303.7). AC acknowledges hospitality of the Max Planck Institute for Physics in Munich where part of this work was carried out.
The work of PC, GL and 
AM is partially supported by the Italian Istituto Nazionale di Fisica Nucleare (INFN) through the ``Theoretical Astroparticle Physics'' project
and by the research grant number 2017W4HA7S
``NAT-NET: Neutrino and Astroparticle Theory Network'' under the program
PRIN 2017 funded by the Italian Ministero dell'Universit\`a e della
Ricerca (MUR).
The work of EV was supported in part by the U.S. Department of Energy (DOE) Grant No. DE-SC0009937.
KK is supported by the Grant-in-Aid for the Scientific Research of Japan Society for the Promotion of Science (JSPS) KAKENHI Grant Number (JP17H01130, JP17H06364, JP17H06357), and by the Central Research Institute of Explosive Stellar Phenomena (REISEP) of Fukuoka University and the associated project (No. 207002).
\appendix
\section*{Appendix: Plasmon dispersions and renormalization factors}
\label{Appendix}

In this Appendix we recollect the dispersion relations for the electromagnetic excitations in a QED plasma~\cite{Raffelt:1996wa}. For an isotropic medium, in the Lorenz gauge, these dispersion relations for the longitudinal and transverse modes read
\begin{subequations}
\begin{align}
	&\omega^2 - v_{*}^2 k^2 = \omega_0^L(k) \equiv \omega_p^2(1-G(v_*^2k^2/\omega^2))\, ,\\
	&\omega^2 - k^2 = \omega_0^T(k) \equiv \omega_p^2(1+1/2\, G(v_*^2k^2/\omega^2)),
	\label{res_L}
\end{align}
\end{subequations}
where G is an auxiliary function, defined as $G(x)=\frac{3}{x}\left[1-\frac{2x}{3}-\frac{1-x}{2\sqrt{x}}\ln{\left(\frac{1+\sqrt{x}}{1-\sqrt{x}}\right)}\right]$, and $v_*$ is the typical velocity for the electron and positron in the plasma~\cite{Raffelt:1996wa}. For the SN profile we considered, $v_* \gtrsim  0.99$ in all the star. 

The renormalization factors are also needed to evaluate Eq.~\eqref{eq:imPIaxion},
\begin{subequations}
\begin{align}
	&Z_L=\frac{\omega^2}{\omega^2-k^2}\frac{2(\omega^2-v_*^2k^2)}{3\omega_p^2-(\omega^2-v_*^2k^2)}\, ,\\
	&Z_T=\frac{2(\omega^2-v_*^2k^2)}{\omega^2(3\omega_p^2-2(\omega^2-k^2))+(\omega^2 + k^2)(\omega^2 - v_*^2k^2)} \, .
\end{align}
\end{subequations}
One finds that $Z_T$ is always very close to unity, while $Z_L$ can strongly deviate from 1.



\bibliographystyle{bibi}
\bibliography{biblio.bib}

\end{document}